\documentclass{IEEEtran}

\usepackage{graphicx}
\usepackage{alltt}

\usepackage[ruled,vlined]{algorithm2e}
\usepackage{algorithmic}
\usepackage{amsfonts}
\usepackage{amssymb}
\usepackage{graphicx}
\usepackage{url}
\usepackage{subfigure}
\usepackage{epstopdf}
\usepackage{multirow}
\usepackage{subfigure}
\usepackage{ifthen}

\newcommand{\V}{V}

\newcommand{\dimension}{\mathbb{D}}

\newcommand{\C}{\mathbb{C}}
\renewcommand{\P}{\mathbb{P}}

\newcommand{\FB}{{\sc FactorBase~}}
\newcommand{\Y}{Y}

\newcommand{\y}{y}

\newcommand{\TTuple}[1][0.0ex]{\vec{t}\hspace{#1}}

\newcommand{\UTuple}[1][0.0ex]{\vec{u}\hspace{#1}}

\newcommand{\VTuple}{\vec{v}}

\newcommand{\Mrange}[1]{\ifthenelse{\equal{#1}{T}}{\TTuple_m}{\ifthenelse{\equal{#1}{U}}{\UTuple_m}{\ifthenelse{\equal{#1}{V}}{\VTuple_m}{\mbox{UNKNOWN
TERM ID}}}}}
\newcommand{\Prange}[1]{\ifthenelse{\equal{#1}{T}}{\vec{t}_{pa}}{\ifthenelse{\equal{#1}{U}}{\vec{u}_{pa}}{\ifthenelse{\equal{#1}{V}}{\vec{v}_{pa}}{\mbox{UNKNOWN
TERM ID}}}}}

\newcommand{\GroundPrange}[1]{\ifthenelse{\equal{#1}{T}}{\vec{t}_{pa,\grounding'}}{\ifthenelse{\equal{#1}{U}}{\vec{u}_{pa,\grounding'}}{\ifthenelse{\equal{#1}{V}}{\vec{v}_{pa,\grounding'}}{\mbox{UNKNOWN
TERM ID}}}}}

\newcommand{\qcount}{\it{count}}

\newcommand{\cttable}{\it{CT}}
\newcommand{\cptable}{\it{CPT}}

\newcommand{\parameters}{\mathit{\#par}}
\newcommand{\aic}{\mathit{AIC}}

\newcommand{\instantiations}{\mathcal{I}}
\newcommand{\parrvs}{\mathsf{A}}
\newcommand{\gparrvs}{\set{A}}
\newcommand{\parfactor}{\Phi}
\newcommand{\parfactors}{\mathcal{F}}

\newcommand{\term}{\tau}

\newcommand{\grounding}{\gamma}

\newcommand{\attribute}{a} 

\newcommand{\RRV}{par-RV}
\newcommand{\RVD}{{\it VDB }}
\newcommand{\CDB}{{\it CDB }}
\newcommand{\MDB}{\it{MDB }}

\newcommand{\loglikelihood}{\it{loglikelihood}}

\newcommand{\G}{G}

\newcommand{\potential}{\phi}

\newcommand{\D}{\mathcal{D}} 
\renewcommand{\S}{\mathbb{S}} 

\newcommand{\student}{\mathit{Student}}

\newcommand{\prof}{\mathit{Professor}}

\newcommand{\ra}{\mathit{RA}}

\newcommand{\true}{\mathrm{T}}
\newcommand{\false}{\mathrm{F}}
\newcommand{\functor}{f}

\def\set#1{\mathbf{#1}}

\graphicspath{{../../}{figures/}}

\begin{document}

\title{\FB: SQL for Learning A  Multi-Relational Graphical Model}
\author{
Oliver Schulte, Zhensong Qian\\
 Simon Fraser University, Canada\\
\{oschulte,zqian\}@sfu.ca
}

\maketitle  
\begin{abstract} 
We describe \FB, a new SQL-based framework that leverages a relational database management system to support multi-relational model discovery. A multi-relational statistical model provides an integrated analysis of the heterogeneous and interdependent data resources in the database.  We adopt the BayesStore design philosophy: statistical models are stored and managed as first-class citizens inside a database \cite{Wang2008}. Whereas previous systems like BayesStore support multi-relational inference, \FB\ supports multi-relational learning. 
A case study on six benchmark databases evaluates how our system supports a challenging machine learning application, namely learning a first-order Bayesian network model for an entire database. Model learning in this setting has to examine a large number of potential statistical associations across data tables. Our implementation shows how the SQL constructs in \FB\ facilitate the fast, modular, and reliable development of highly scalable model learning systems.

\end{abstract}

\section{Introduction} Data science brings together ideas from different fields for extracting value from large complex datasets. The system described in this paper combines advanced analytics from multi-relational or {\em statistical-relational} machine learning (SRL) with database systems. The power of combining machine learning with database systems has been demonstrated in several systems 
\cite{MADlib_VLDB_2012,MLbase_ICDR_2013,Deshpande2006}.
The novel contribution of \FB\ is supporting machine learning for {\em multi-relational} data, rather than for traditional learning where the  data are represented in a {\em single} table or data matrix.
We discuss new challenges raised by multi-relational model learning compared to single-table learning, and how \FB\ solves them using the resources of SQL (Structured Query Language). The name \FB\ indicates that our system supports learning factors that define a log-linear multi-relational model~\cite{Kimmig2015}. Supported new database services include constructing, storing, and transforming complex statistical objects, such as factor-tables, cross-table sufficient statistics, parameter estimates, and model selection scores.

Multi-relational data have a complex structure, that integrates heterogeneous information about different types of entities (customers, products, factories etc.) and different types of relationships among these entities. 
A statistical-relational model provides an integrated statistical analysis of the heterogeneous and interdependent complex data resources maintained by the database system. 
Statistical-relational models have achieved state-of-the-art performance in a number of application domains, such as natural language processing, ontology matching, information extraction, entity resolution, link-based clustering, query optimization, etc. 
\cite{Domingos2009,Niu2011,Getoor2001}.
Database researchers have noted the usefulness of statistical-relational models for knowledge discovery and for representing uncertainty in databases 
\cite{Singh2013,Wang2008}.
They have developed a system architecture where statistical models are stored as first-class citizens {\em inside a database.} The goal is to seamlessly integrate query processing and statistical-relational inference. 
These systems focus  on inference {\em given} a statistical-relational model, not on {\em learning} the model from the data stored in the RDBMS.
\FB\ complements the in-database probabilistic inference systems by providing an in-database probabilistic model learning system.

\subsection{Evaluation} We evaluate our approach on six benchmark databases. For each benchmark database, the system applies a state-of-the-art SRL algorithm to construct a statistical-relational model.
Our experiments show that \FB\ pushes the scalability boundary: Learning scales to databases with over $10^6$ records, compared to less than $10^5$ for previous systems. At the same time it is able to discover more complex cross-table correlations than previous SRL systems. We report experiments that focus on two key services for an SRL client: (1) Computing and caching sufficient statistics, (2) computing model predictions on test instances. For the largest benchmark database, our system handles 15M sufficient statistics. 
SQL facilitates block-prediction for a set of test instances, which leads to a 10 to 100-fold speedup compared to a simple loop over test instances.

\subsection{Contributions}
\FB  is the first system that leverages relational query processing for learning a multi-relational log-linear graphical model. Whereas the in-database design philosophy has been previously used for multi-relational inference, we are the first to adapt it for multi-relational model structure learning. Pushing the graphical model inside the database 
allows us to {\em use SQL as a high-level scripting language for SRL}, with the following advantages.

\begin{enumerate}
\item Extensibility and modularity, which support rapid prototyping. SRL algorithm development can focus on statistical issues and rely on a RDBMS for data access and model management.
\item Increased scalability, in terms of both the size and the complexity of the statistical objects that can be handled.
\item Generality and portability: standardized database operations support ``out-of-the-box'' learning with a minimal need for user configuration.
\end{enumerate}
\subsection{Paper Organization}
We provide an overview of the system components and flow. For each component, we describe how the component is constructed and managed inside an RDBMS using SQL scripts and the SQL view mechanism. We show how the system manages sufficient statistics and test instance predictions. The evaluation section demonstrates the scalability advantages of in-database processing. The intersection of machine learning and database management has become a densely researched area, so we end with an extensive discussion of related work.

\section{Background on Statistical-Relational Learning} We review enough background from statistical-relational models and structure learning to motivate our system design. The extensive survey by Kimmig {\em et al.} \cite{Kimmig2015} provides further details. The survey shows that SRL models can be viewed as log-linear models based on par-factors, as follows.

\subsection{Log-linear Template Models for Relational Data} \label{sec:log-linear}

Par-factor stands for ``parametrized factor''. A par factor represents an interaction among parametrized random variables, or par-RVs for short. 
We employ the following notation 
for par-RVs \cite[2.2.5]{Kimmig2015}.
	Constants are expressed in lower-case, e.g. $\it{joe}$, and are used to represent entities. A type is associated with each entity, e.g. $\it{joe}$ is a person. 
A first-order variable is also typed, e.g. $\it{Person}$ denotes some member of the class of persons. A functor maps a tuples of entities to a value. We assume that the range of possible values is finite.  An {\em atom} is an expression of the form $r(\term_{1},\ldots,\term_{a})$ where each $\term_{i}$ is either a constant or a first-order variable. If all of $\term_{1},\ldots,\term_{a}$ are constants, $r(\term_{1},\ldots,\term_{a})$ is a {\em ground atom} or random variable (RV), otherwise a {\em first-order atom} or a \textbf{par-RV}. A par-RV is instantiated to an RV by grounding, i.e. substituting a constant of the appropriate domain for each first-order variable. 

A \textbf{par-factor} is a pair $\parfactor = (\parrvs,\potential)$, where $\parrvs$ is a set of par-RVs, and $\potential$ is a function from the values of the par-RVs to the non-negative real numbers.\footnote{A par-factor can also include constraints on possible groundings.} Intuitively, a grounding of a par-factor represents a set of random variables that interact with each other locally. SRL models use {\em parameter tying}, meaning that if two groundings of the same par-factor are assigned the same values, they return the same factor value. A set of parfactors $\parfactors$ defines a joint probability distribution over the ground par-RVs as follows. Let $\instantiations(\parfactor_{i})$ denote the set of {\em all} ground par-RVs in par-factor $\parfactor_{i}$. Let $\set{x}$ be a joint assignment of values to all ground random variables. Notice that this assignment determines the values of all ground atoms. An assignment $\set{X}=\set{x}$ is therefore {\em equivalent to a single database instance}.
The probability of a database instance is given by the log-linear equation \cite[Eq.7]{Kimmig2015}:
\begin{equation} \label{eq:parfactor}
P(\set{X}=\set{x}) = \frac{1}{Z} \prod_{\parfactor_{i} \in \parfactors} \prod_{\gparrvs \in \instantiations(\parfactor_{i})} 
\potential_{i}(\set{x}_{\gparrvs}) 
\end{equation}
where $\set{x}_{\gparrvs}$ represents the values of those variables in $\gparrvs$ that are necessary to compute $\potential_{i}$. 
Equation~\ref{eq:parfactor} can be read as follows.

\begin{enumerate}
\item Instantiate all parfactors with all possible constants. 
\item For each ground par-factor, apply the values specified by the joint assignment  $\set{X}=\set{x}$, and compute the corresponding factor term.
\item Multiply all the factors together, and normalize the product.  
\end{enumerate}

Typically the number of ground par-RVs $\instantiations(\parfactor_{i})$ will be very large.  
Equation~\ref{eq:parfactor} can be evaluated, without enumerating the ground par-factors, 
as follows. 

\begin{enumerate}
\item For each par-factor, for each possible assignment of values, find the number of ground factors with that assignment of values.
\item Raise the factor value for that assignment to the number of instantiating factors.
\item Multiply the exponentiated factor values together, and normalize the product. 
\end{enumerate}

The number 2) of ground factors with the same assignment of values is known as a \textbf{sufficient statistic} of the log-linear model. 

\begin{figure}[htbp] 
 \centering
\resizebox{0.45\textwidth}{!}{
 \includegraphics{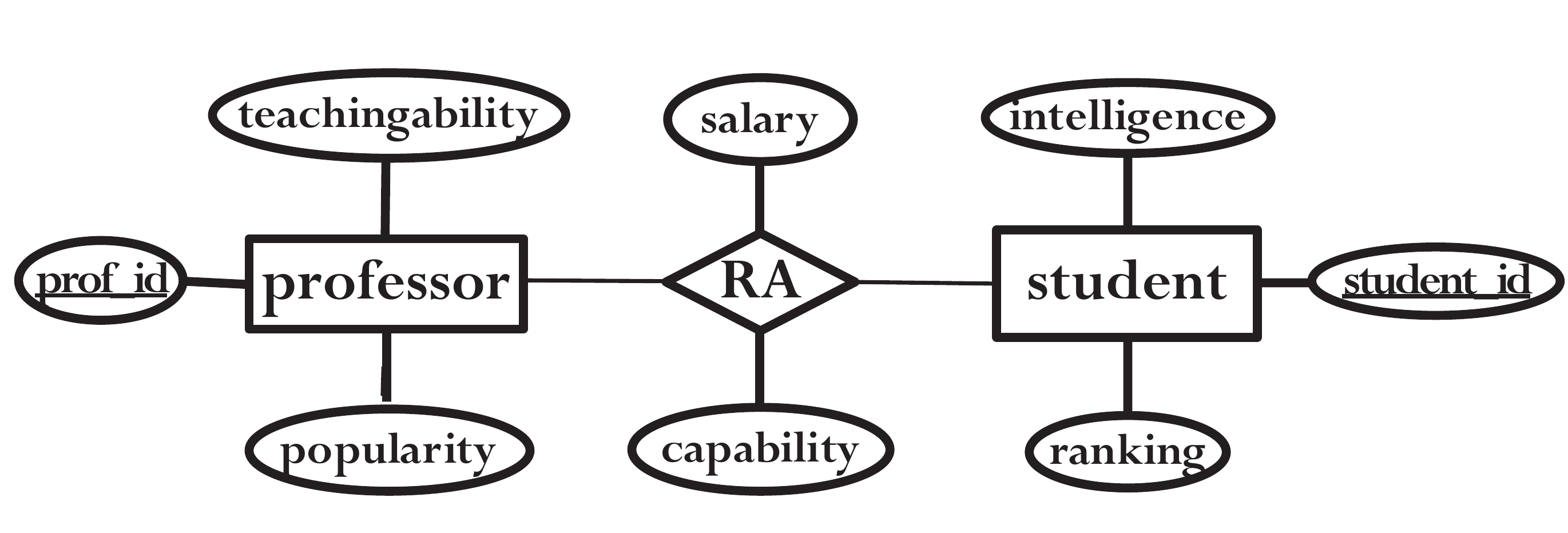} 
 } 
\caption{A relational ER Design for a university domain.}
 \label{fig:university-schema}
\end{figure}
\begin{figure}[htbp] 
 \centering
\resizebox{0.5\textwidth}{!}{
 \includegraphics[width=0.5\textwidth]{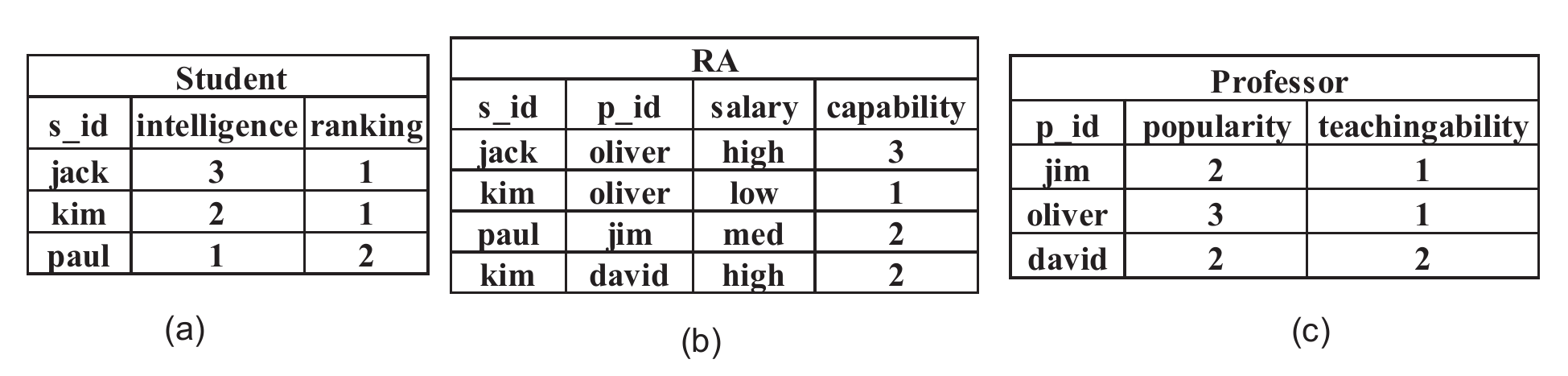} 
} 
\caption{Database Table Instances: (a) $\student$, (b) $\ra$, (c) $\prof$. }
 \label{fig:instance}
\end{figure}
\begin{figure}[htbp] 
 \centering
\resizebox{0.4\textwidth}{!}{
 \includegraphics{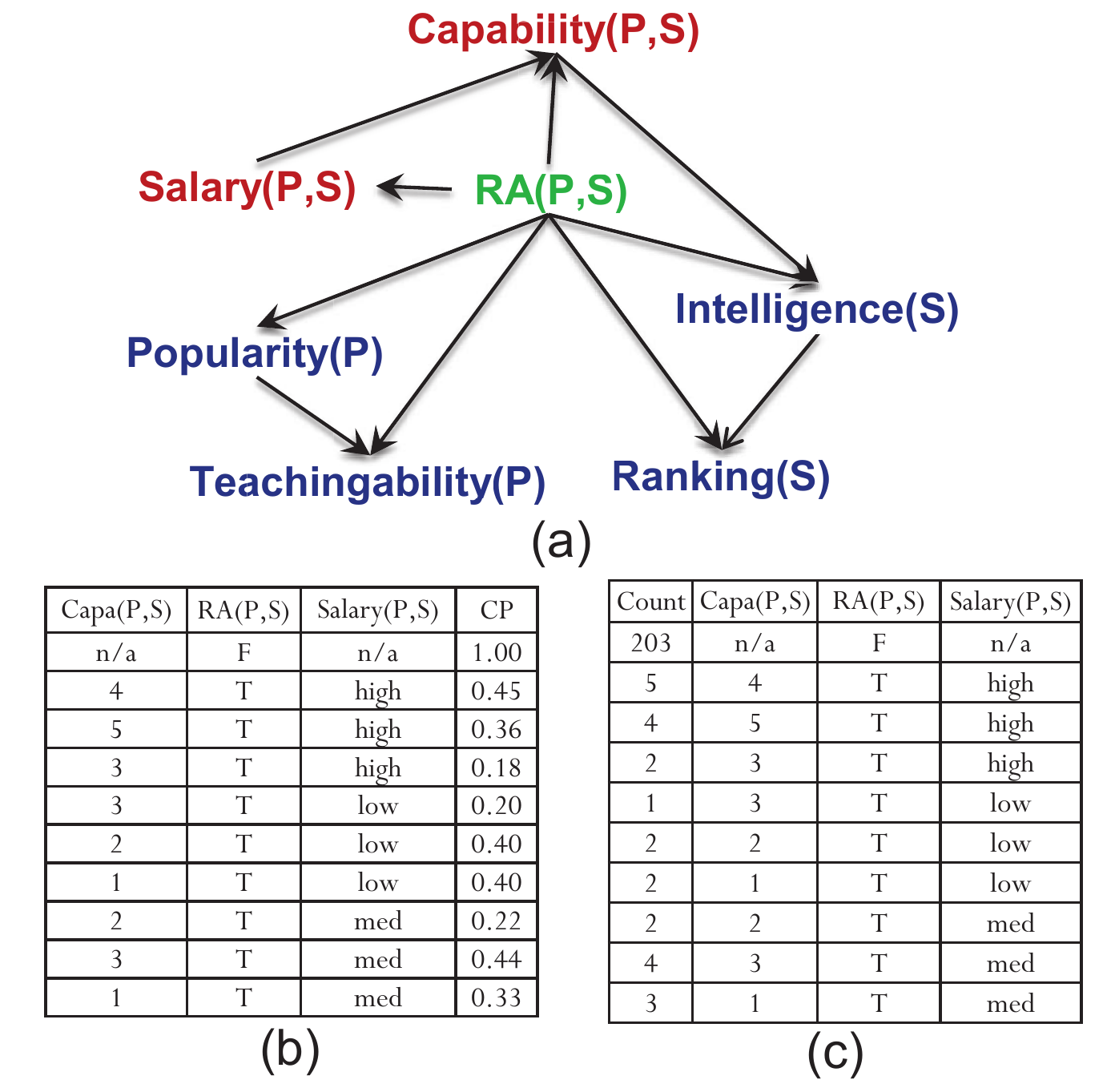} 
} 
\caption{(a) Bayesian network for the University domain. We omit the $\it{Registered}$ relationship for simplicity. The network was learned from the University dataset \cite{bib:bbsite}.
(b) Conditional Probability table $Capability(\P,\S)\_\cptable$, for the node $Capability(\P,\S)$. Only value combinations that occur in the data are shown. This is an example of a factor table. (c) Contingency Table $Capability(\P,\S)\_\cttable$ for the node $Capability(\P,\S)$ and its parents. Both CP and CT tables are stored in an RDBMS.}
 \label{fig:pbn}
\label{fig:ct-cp-table}
\end{figure}

\subsection{Examples} \label{sec:examples} SRL has developed a number of formalisms for describing par-factors \cite{Kimmig2015}. 
First-order probabilistic graphical models
are popular both within SRL and the database community \cite{Kimmig2015,Wang2008}. The model structure is defined by edges connecting par-RVs. For instance, a \textbf{parametrized Bayesian network structure} is a directed acyclic graph whose nodes are par-RVs.  Figure \ref{fig:pbn} shows a Bayesian network for a University domain. We use the university example as a toy running example throughout the paper. The schema for the university domain is given in Figure~\ref{fig:university-schema}. This schema features only one relationship for simplicity; \FB\ learns a model for any number of relationships. While we describe \FB\ abstractly in terms of par-factors, for concreteness we illustrate it using Bayesian networks. The system takes as input a database instance like that shown in Figure~\ref{fig:instance}, and produces as output a graphical model like that shown in Figure~\ref{fig:pbn}.

A par-factor in a Bayesian network is associated with a \textbf{family} of nodes \cite[Sec.2.2.1]{Kimmig2015}. A family of nodes comprises a child node and all of its parents. For example, in the BN of Figure~\ref{fig:pbn}, one of the par-factors is associated with the par-RV set $\parrvs=\{\it{Capability}(\P,\S),\it{Salary}(\P,\S),\it{RA}(\P,\S)\}$. For the database instance of Figure~\ref{fig:instance}, there are $3\times3=9$ possible factors associated with this par-RV, corresponding to the Cartesian product of 3 professors and 3 students. The value of the factor $\phi$ is a function from an assignment of family node values to a non-negative real number. {\em In a Bayesian network, the factor value represents the conditional probability of the child node value given its parent node values.} These conditional probabilities are typically stored in a table as shown in Figure~\ref{fig:pbn}(b). This table represents therefore the function $\phi$ associated with the family par-factor. Assuming that all par-RVs have finite domains, a factor can always be represented by a \textbf{factor table} of the form Figure~\ref{fig:pbn}(b): there is a column for each par-RV in the factor, each row specifies a joint assignment of values to a par-RV, and the factor column gives the value of the factor for that assignment (cf. \cite[Sec.2.2.1]{Kimmig2015}).

The sufficient statistics for the $\it{Capability}(\P,\S)$ family can be represented in a contingency table as shown in Figure~\ref{fig:pbn}(c). For example, the first row of the contingency table indicates that the conjunction \\$\it{Capability}(\P,\S)=n/a,\it{Salary}(\P,\S)=n/a,\it{RA}(\P,\S) =\false$ is instantiated 203 times in the University database (publicly available at~\cite{bib:bbsite}). This means that for 203 professor-student pairs, the professor did not employ the student as an RA (and therefore the salary and capability of this RA relationship is undefined or $n/a$).

\subsection{SRL Structure Learning}

\begin{figure*}[t]
\begin{center}
\resizebox{1\textwidth}{!}{
\includegraphics
{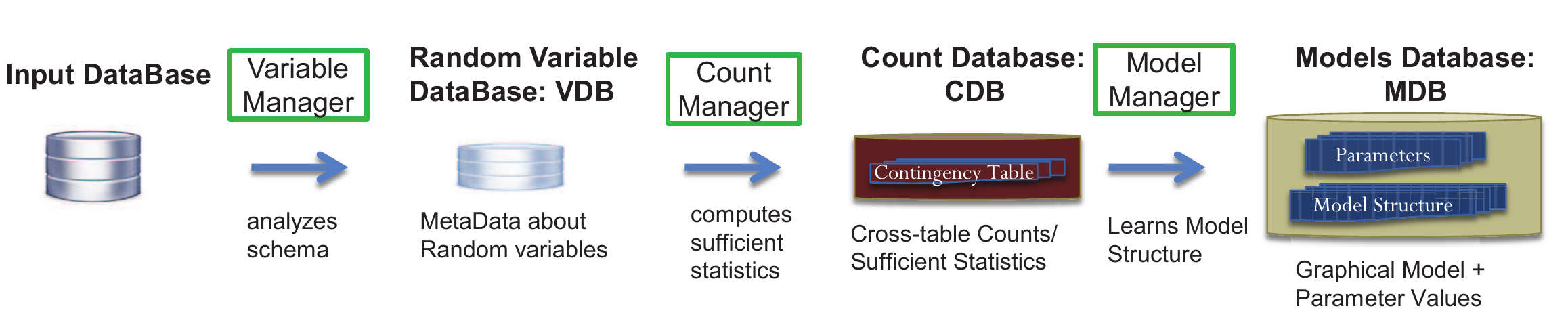}
}
\caption{System Flow. All statistical objects are stored as first-class citizens in a DBMS. Objects on the left of an arrow are utilized for constructing objects on the right. Statistical objects are constructed and managed by different modules, shown as boxes. 
\label{fig:architecture}}
\end{center}
\end{figure*}

Algorithm~\ref{alg:learning} shows the generic format of a statistical-relational structure learning algorithm (adapted from 
\cite{Kimmig2015}
). The instantiation of procedures in lines 2, 3, 5 and 8 determines the exact behavior of a specific learning algorithm. The structure algorithm carries out a local search in the hypothesis space of graphical relational models. A set of candidates is generated based on the current model (line 3), typically using a search heuristic. For each candidate model, parameter values are estimated that maximize a model selection score function chosen by the  user (line 5). A model selection score is computed for each model given the parameter values, and the best-scoring candidate model is selected (line 7). 
We next discuss our system design and how it supports model discovery algorithms that follow the outline of Algorithm~\ref{alg:learning}. Figure~\ref{fig:architecture} outlines the system components and dependencies among them.

\begin{algorithm}[htbp]

\KwIn{Hypothesis space $ \mathcal H$ (describing graphical models), training data $\mathcal D$ (assignments to random variables), scoring function score ($\cdot$, $\mathcal D$)}
\KwOut{A graph structure $G$ representing par-factors.}
\begin{algorithmic}[1]
\STATE $G \leftarrow 	\emptyset$
\WHILE{ \textsc{continue}($G$, 
$\mathcal H$, score ($\cdot$, $\mathcal D$) )} 
	\STATE{ $\mathcal R \leftarrow $ \textsc{refine}C\textsc{andidates}($G, \mathcal H$)} 
 	\FORALL{$R \in \mathcal R$} 
		\STATE{$R \leftarrow$ \textsc{learn}P\textsc{arameters}($R$,score ($\cdot$, $\mathcal D$))} 
    \ENDFOR
	\STATE 	$G \leftarrow$ argmax$_{G^{\prime} \in \mathcal R\cup \{G\}}$ score($G^{\prime} $, $\mathcal D$)
\ENDWHILE
\STATE \Return $G$
\end{algorithmic}
\label{alg:learning}
\caption{Structure learning algorithm 
}
\end{algorithm}

\section{The Random Variable Database} 

Statistical-relational learning requires various metadata about the par-RVs in the model. These include the following. 

\begin{LaTeXdescription}
\item[Domain] the set of possible values of the par-RV.
\item[Types] Pointers to the first-order variables
in the par-RV. 
\item[Data Link] Pointers to the table and/or column in the input database associated with the par-RV. 
\end{LaTeXdescription}

The metadata must be machine-readable. Following the in-database design philosophy, we store the metadata in tables so that an SRL algorithm can query it using SQL. The schema analyzer uses an SQL script that queries key constraints in the system catalog database and {\em automatically} converts them into metadata stored in the random variable database $VDB$. In contrast, existing SRL systems require users to specify information about par-RVs and associated types. 
Thus \FB\  utilizes the data description resources of SQL to facilitate the ``setup task'' for relational learning \cite{Walker2010}. Due to space constraints, we do not go into the details of the schema analyzer. Instead, we illustrate the general principles with the ER diagram of the University domain (Figure~\ref{fig:university-schema}). The Appendix  provides a full description with MySQL script. 

\begin{figure}[htbp] 
 \centering
\resizebox{0.4\textwidth}{!}{
 \includegraphics[width=0.5\textwidth]{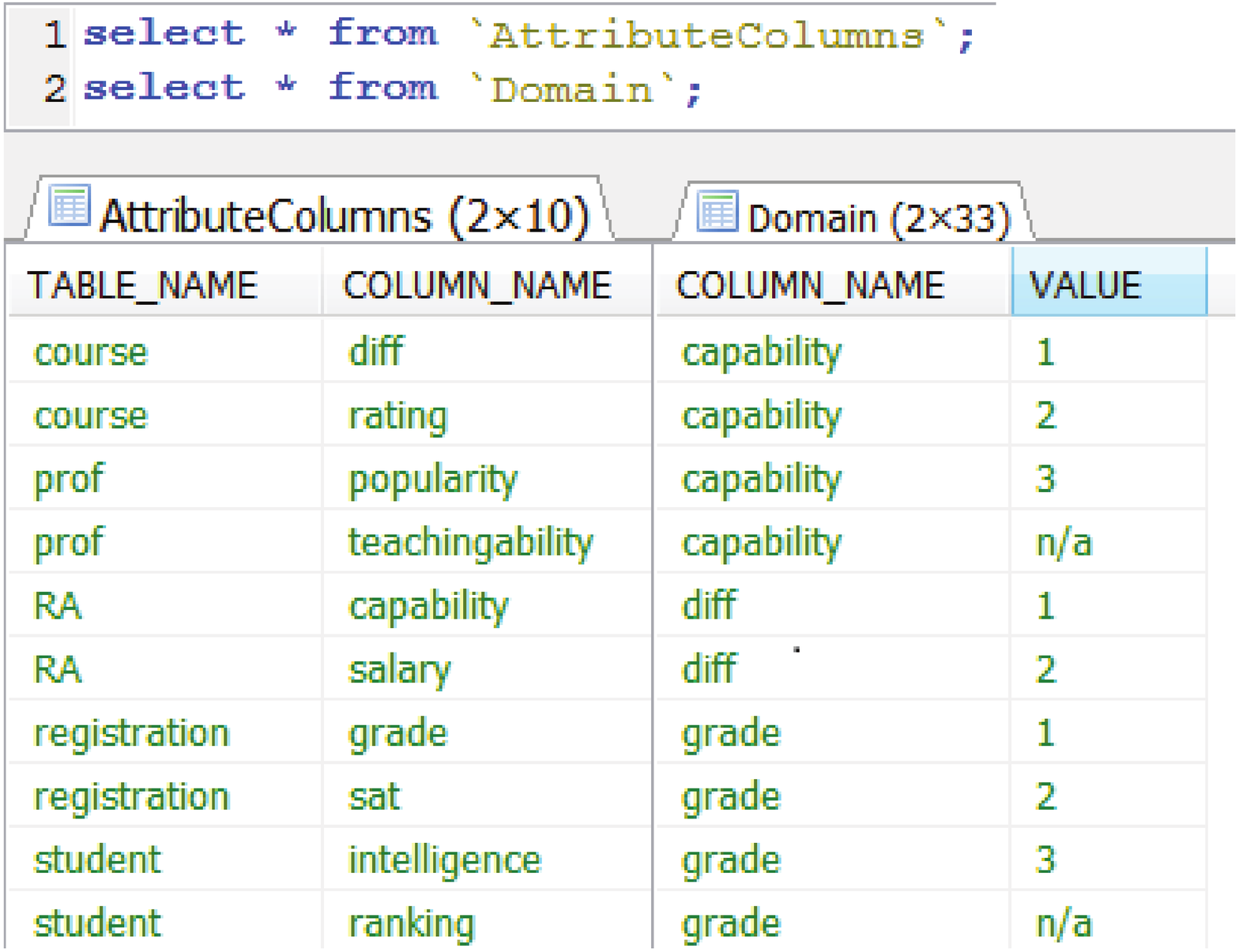} 
} 
\caption{ The metadata about attributes represented in \RVD database tables.  Left: The table $\it{AttributeColumns}$ specifies which tables and columns contain the functor values observed in the data. The column name is also the functor ID. Right: The table $\it{Domain}$ lists the domain for each functor.
}
 \label{fig:attributes}
\end{figure}

The translation of an ER diagram into a set of functors converts each element of the diagram into a functor, except for entity sets and key fields~\cite{Heckerman+al:SRL07}. Table~\ref{table:translation} illustrates this translation. In terms of database tables, attribute par-RVs correspond to {\em columns}. Relationship par-RVs correspond to {\em tables}, not columns. Including a relationship par-RV in a statistical model allows the model to represent uncertainty about whether or not a relationship exists between two entities~\cite{Kimmig2015}. The values of descriptive attributes of relationships are undefined for entities that are not related. We represent this by introducing a new constant $\it{n/a}$ in the domain of a relationship attribute~\cite{Milch2005}; see Figure~\ref{fig:attributes} (right). Table~\ref{table:vdb-schema} shows the schema for some of the tables that store metadata for 
each relationship par-RV, as follows. par-RV and FO-Var are custom types.

\begin{LaTeXdescription}
\item[Relationship] The associated input data table.
\item[Relationship\_Attributes] Descriptive attributes associated with the relationship and with the entities involved.
\item[Relationship\_FOVariables] The first-order variables contained in each relationship par-RV.\footnote{The schema assumes that all relationships are binary.}
\end{LaTeXdescription}

\begin{table}[btp]
\caption{Translation from ER Diagram to Par-RVs}
 \centering
\resizebox{0.5\textwidth}{!}{
\begin{tabular}[c]{|l|l|l|}\hline
 ER Diagram &  Example &par-RV equivalent \\\hline
Entity Set  &Student, Course & $\S, \C$ \\\hline
 Relationship Set &RA & RA($\P$,$\S$) \\\hline
Entity Attributes  &intelligence, ranking & Intelligence($\S$), Ranking($\S$) \\\hline
Relationship Attributes  &capability, salary &Capability($\P, \S$), Salary($\P, \S$) \\\hline
\end{tabular}
}
 
 \label{table:translation}
\end{table}

\begin{table}[hbtp]
\caption{Selected Tables In the Variable Database Schema.}
 \centering
\begin{tabular}{|l|p{6cm}|}
\hline
    {Table Name}&  {Column Names} \\\hline
{Relationship} & {RVarID: par-RV,  TABLE\_NAME: string} \\\hline
    {Relationship\_Attributes}&  {RVarID: par-RV, AVarID: par-RV,  FO-ID: FO-Var} \\\hline
    {Relationship\_FOvariables} & {RVarID: par-RV, FO-ID: FO-Var,  TABLE\_NAME: string}\\\hline
        \end{tabular}%
 \label{table:vdb-schema}
\end{table}

While we have described constructing the variable database for an ER model, different structured data models can be represented by an appropriate first-order logic vocabulary \cite{Kimmig2015}, that is, an appropriate choice of functors. For example, in a star schema, facts can be represented in the form $\functor(\dimension_{1},\ldots,\dimension_{k})$, where the first-order variable $\dimension_{i}$ ranges over the primary key of dimension table $i$. Attributes of dimension $i$ can be represented by a unary functor $\attribute(\dimension_{i})$. \FB can perform structure learning for different data models after the corresponding data format has been translated into the \RVD format.

\section{The Count Manager}

The \textbf{count database} \CDB stores a set of  {\em contingency tables}. Contingency tables represent sufficient statistics as follows~\cite{Moore1998}. 
Consider a fixed list of par-RVs.
A \textbf{query} is a set of $(variable = value)$ pairs where each value is of a valid type for the variable. 
The \textbf{result set} of a query in a database $\D$ is the set of instantiations of the logical variables such that the query evaluates as true in $\D$.
For example, in the database of Figure~\ref{fig:instance} the result set for the query\\ 
$\it{RA}(\P,\S) = \true$, $\it{Capability}(\P,\S) = 3$, $\it{Salary}(\P,\S) = \it{high} $\\ is
the singleton $\{\langle \it{jack}, \it{oliver}\rangle\}$. 
The \textbf{count} of a query is the cardinality of its result set. 

Every set of par-RVs $\set{V} \equiv \{\V_{1},\ldots,\V_{n} \}$ has an associated \textbf{contingency table} ($\cttable$) denoted by $\cttable(\set{V})$. 
This is a table with a row for each of the possible assignments of values to the variables in $\set{V}$, and a special integer column called $\qcount$. 
The value of the $\qcount$ column in a row 
corresponding to $V_{1} = v_{1},\ldots,V_{n} = v_{n}$ records the count of the 
corresponding query. 
Figure~\ref{fig:pbn}(b) shows a contingency table for the par-RVs $\it{RA}(\P,\S)$, $\it{Capability}(\P,\S)$, $\it{Salary}(\P,\S)$. The \textbf{contingency table problem} is to compute a contingency table for par-RVs $\set{V} $ and an input database $\D$.

{\em SQL Implementation With Metaqueries.}
We describe how the contingency table problem can be solved using SQL. 
This is relatively easy for a {\em fixed} set of par-RVs; the challenge is a general construction that works for different sets of par-RVs. For a fixed set, a  contingency table can be computed by an SQL count(*) query of the form 
\begin{alltt}
CREATE VIEW CT-table(<VARIABLE-LIST>) AS
SELECT COUNT(*) AS count, <VARIABLE-LIST>
FROM TABLE-LIST
GROUP BY VARIABLE-LIST
WHERE <Join-Conditions>
\end{alltt}

\FB\ uses SQL itself to construct the count-conjunction query. We refer to this construction as an SQL \textbf{metaquery}. We represent a count(*) query in 
four kinds of tables: the Select, From, Where and Group By tables. Each of these tables lists the entries in the corresponding count(*) query part.
Given the four metaquery tables, the corresponding SQL count(*) query can be easily constructed and executed in an application to construct the contingency table.
Given a list of par-RVs as input, the metaquery tables are constructed as follows
from the metadata in the database $\RVD$.

\begin{LaTeXdescription}
\item[FROM LIST] Find the tables referenced by the \RRV's. A \RRV ~references the entity tables associated with its first-order variables (see VDB.Relationship\_FOvariables). Relational \RRV's also reference the associated relationship table (see VDB.Relationship). 
\item[WHERE LIST] Add join conditions on the matching primary keys of the referenced tables in the WHERE clause. The primary key columns are recorded in VDB. 
\item[SELECT LIST] For each attribute \RRV, find the corresponding column name in the original database (see VDB.AttributeColumns). Rename the column with the ID of the \RRV. Add a $\qcount$ column.
\item[GROUP BY LIST] The entries of the Group By table are the same as in the Select table without the $\qcount$ column.
\end{LaTeXdescription}


\begin{figure}[htb]
\begin{center}
\resizebox{0.45\textwidth}{!}{
\includegraphics[width=0.45\textwidth]{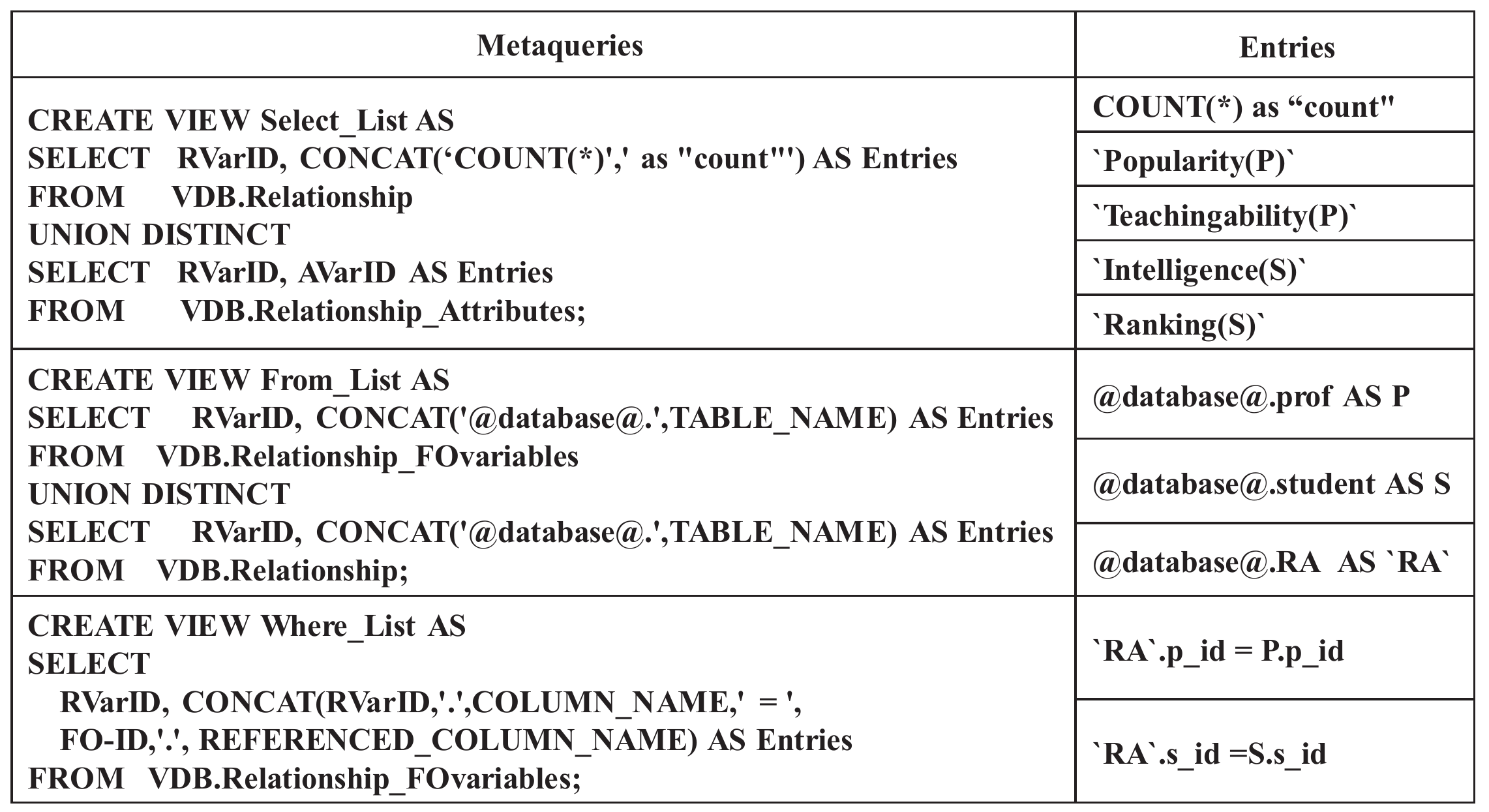}
}
\caption{Example of metaquery results based on university database and the par-RV metadata (Table \ref{table:vdb-schema}).
~\label{fig:meta-query}}
\end{center}
\end{figure}

Figure~\ref{fig:meta-query} shows an example of a metaquery for the university database. This metaquery defines a view that in turn defines a contingency table for the random variable list associated with the relationship table $\ra$. This list includes the entity attributes of professors and of students, as well as the relationship attributes of the $\ra$ relationship. 
The Bayesian network of Figure~\ref{fig:pbn} was learned from this contingency table. 
The  contingency table defined by the metaquery of Figure~\ref{fig:meta-query} contains only rows where the value of $\ra$ is true. The M\"obius Virtual Join~\cite{Qian2014a} can be used to extend this contingency table to include counts for when $\ra$ is false, like the table shown in Figure~\ref{fig:pbn}(c).

\section{The Model Manager}

\begin{figure}[htbp]
\begin{center}
\resizebox{0.4\textwidth}{!}{
\includegraphics
{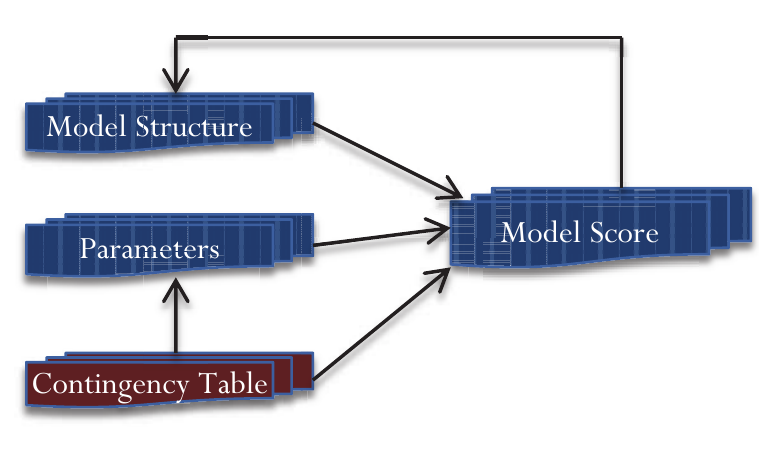}
}
\caption{Dependencies Among Key Components of the Model  Manager. 
\label{fig:learning}}
\end{center}
\end{figure}

The Model Manager provides two key services for statistical-relational structure learning: 1) Estimating and storing parameter values (line 5 of Algorithm~\ref{alg:learning}). 2) Computing one or more model selection scores (line 7 of Algorithm~\ref{alg:learning}).  \FB\ uses a {\em store+score} design for these services, which is illustrated in 
Figure~\ref{fig:learning}. 
A \textbf{model structure table} represents a candidate model. When a candidate model structure is inserted, a view uses the sufficient statistics from a contingency table to compute a table of parameter values. Another view uses the parameter values and sufficient statistics together to compute the score for the candidate model. 

\subsection{The \MDB Schema}

The relational schema for the Models Database is shown in Table~\ref{table:mdb-schema}. The @par-RVID@ parameter refers to the ID  of a par-RV, for instance $\it{Capability}(\P,\S)$.
The model manager stores a set of factor tables (cf. Section~\ref{sec:examples}). In a graphical model, each factor is defined by the local topology of the model template graph. For concreteness, we illustrate how factor tables can be represented  for Bayesian networks. The graph structure can be stored straightforwardly in a database table $\it{BayesNet}$ whose columns are $\it{child}$ and $\it{parent}$. The table entries are the IDs of par-RVs. 
For each node, the $\MDB$ manages a conditional probability table. This is a factor table that represents the factor associated with the node's family (see Figure~\ref{fig:pbn}(b)).
In a Bayesian network, model selection scores are decomposable. This means that there is a local score associated with each family, such that the total score for the BN model is the sum of the local scores. For each family, the local score is stored in the $\it{Scores}$ table indexed by the family's child node.

\begin{table}[tbp]
\caption{The main tables in the Models Database $\MDB$. For a Bayesian network, the $\MDB$ stores its structure, parameter estimates, and model selection scores.}
 \centering
 \begin{tabular}
[c]{|l|}\hline
BayesNet(\underline{child:par-RV,parent:par-RV})\\
@par-RVID@\_CPT(\underline{@par-RVID@:par-RV,$\mbox{parent}_{1}$:par-RV,$\ldots,\mbox{parent}_{k}$:par-RV},cp:real)\\ 
Scores(\underline{child:par-RV},loglikelihood:real,\#par:int,aic:real)\\
\hline
\end{tabular}
\label{table:mdb-schema}
\end{table}

\subsection{Parameter Manager} \label{sec:parameters}

Deriving predictions from a model requires estimating values for its parameters.  Maximizing the data likelihood is the basic parameter estimation method for Bayesian networks. The maximum likelihood estimates equal the observed frequency of a child value given its parent values.

{\em SQL Implementation With Natural Join.} Given the sufficient statistics in a contingency table, a conditional probability table containing the maximum likelihood estimates can be computed by aggregation using SQL as in the example below. 

\begin{alltt}
SELECT count/temp.parent\_count as CP, 
capability(P,S), RA(P,S), salary(P,S) 
FROM capability(P,S)\_CT 
NATURAL JOIN  
(SELECT sum(Count) as parent\_count, 
RA(P,S), salary(P,S) 
FROM capability(P,S)\_CT  
GROUP BY  RA(P,S), salary(P,S) ) as temp
\end{alltt}

\subsection{Model Score Computation} \label{sec:model-score}
A typical model selection approach is to maximize the likelihood of the data, balanced by a penalty term. For instance, the Akaike Information Criterion (AIC) is defined as follows 
\[
\mathit{AIC}(\G,\D) \equiv ln(P_{\widehat{G}}(\D)) - \parameters(\G) \]
where $\widehat{G}$ is the BN $\G$ with its parameters instantiated to be the maximum likelihood estimates given the database $\D$, and $\parameters(\G)$ is the number of free parameters in the structure $\G$. 
The number of free parameters for a node is the product of (the possible values for the parent nodes) $\times$ (the number of the possible values for the child node -1). Given the likelihood and the number of parameters, the AIC column is computed as $\aic = \loglikelihood - \parameters$. 
 Model selection scores other than AIC can be computed in a similar way given the model likelihood and number of parameters.

\subsubsection{Parameter Number Computation} To determine the number of parameters of the child node @parVar-ID@, the number of possible child and parent values can be found from the $\it{\RVD.Domain}$ table in the Random Variable Database.

\subsubsection{Likelihood Computation} As explained in Section~\ref{sec:log-linear}, the log-likelihood can be computed by multiplying the instantiation counts of a factor by its value. Assuming that instantiation counts are represented in a contingency table and factor values in a factor table, this multiplication can be elegantly performed using the Natural Join operator. For instance, the log-likelihood score associated with the $Capability(\P,\S)$ family is given by the SQL query below. The aggregate computation in this short query illustrates how well SQL constructs support complex computations with structured objects. 

\begin{alltt}
SELECT Capability(P,S),  SUM
(MDB.Capability(P,S)\_CPT.cp * \\CDB.Capability(P,S)\_CT.count) 
AS loglikelihood
FROM MDB.Capability(P,S)\_CPT 
NATURAL JOIN CDB.Capability(P,S)\_CT
\end{alltt}

It is possible to extend the model manager to handle the multi-net structure learning method of the learn-and-join algorithm~\cite{Schulte2012}. The algorithm learns multiple Bayesian networks and propagates edges among them. The $\MDB$ schema is easily extended to store multiple Bayesian networks in a single table. The edge propagation can be executed by the RDBMS using the view mechanism. For more details, please see~\cite{bib:bbsite}. 

This completes our description of how the modules of \FB\ are implemented using SQL. We next show how these modules support a key learning task: computing the predictions of an SRL model on a test instance.

\section{Test Set Predictions} Computing probabilities over the label of a test instance is important for several tasks. 1) Classifying the test instance, which is one of the main applications of a machine learning system for end users. 2) Comparing the class labels predicted against true class labels is a key step in several approaches to model scoring \cite{Kimmig2015}. 3) Evaluating the accuracy of a machine learning algorithm by the train-and-test paradigm, where the system is provided a training set for learning and then we test its predictions on unseen test cases. %
We first discuss how to compute a prediction for a single test case, then how to compute an overall prediction score for a set of test cases. Class probabilities can be derived from Equation~\ref{eq:parfactor} as follows \cite[Sec.2.2.2]{Kimmig2015}. Let $\Y$ denote a ground par-RV to be classified, which we refer to as the \textbf{target variable}. For example, a ground atom may be $\it{Intelligence}(jack)$. In this example, we refer to jack as the \textbf{target entity}. Write $\set{X}_{-\Y}$ for a database instance that specifies the values of all ground par-RVs, except for the target, which are used to predict the target node. Let $[\set{X}_{-\Y},\y]$ denote the completed database instance where the target node is assigned value $\y$. The log-linear model uses the likelihood $P([\set{X}_{-\Y},\y])$ as the joint score of the label and the predictive features. The conditional probability is proportional to this score:
\begin{equation} \label{eq:classify}
P(\y|\set{X_{-\Y}}) \propto P([\set{X}_{-\Y},\y])
\end{equation}
where the joint distribution on the right-hand side is defined by Equation~\ref{eq:parfactor}, and the scores of the possible class labels need to be normalized to define  conditional probabilities.

{\em SQL Implementation.} 
The obvious approach to computing the log-linear score would be to use the likelihood computation of Section~\ref{sec:model-score} for the entire database.
This is inefficient because only instance counts that involve the target entity change the classification probability. 
This means that we need only consider query instantiations that match the appropriate logical variable with the target entity (e.g., $\S = jack$).

\begin{figure}[htbp] 
 \centering
\resizebox{0.35\textwidth}{!}{
 \includegraphics{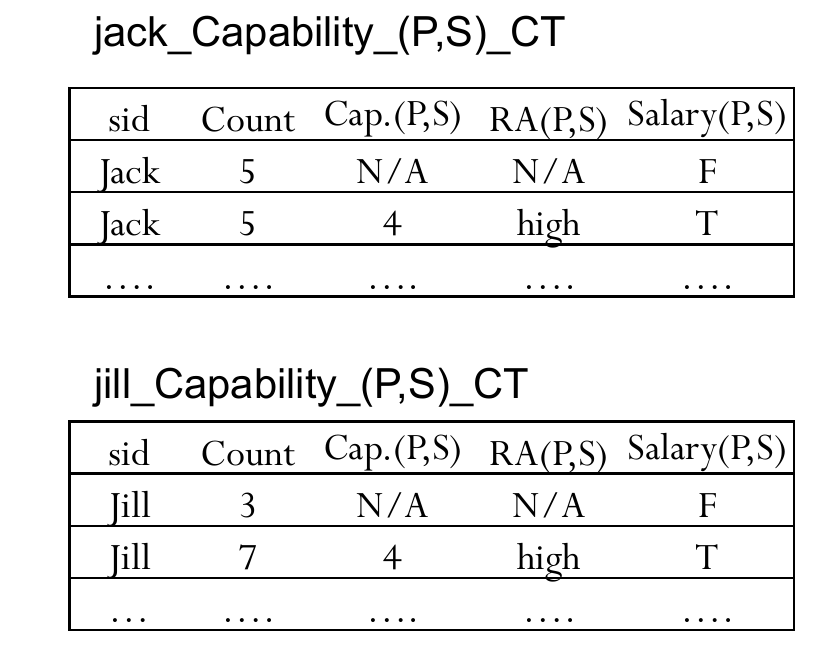} 
} 
\caption{Target contingency tables for target = jack and for target = jill.}
 \label{fig:targetct}
\end{figure}

For a given set of random variables, target entity instantiation counts can be represented in a contingency table that we call the \textbf{target contingency table}. Figure~\ref{fig:targetct} shows the format of a contingency table for target entities jack resp. jill.

\begin{table*}[t]
\caption{SQL queries for computing target contingency tables supporting test set prediction.  $\textless$Attribute-List$\textgreater$ and  $\textless$Key-Equality-List$\textgreater$ are as in Figure~\ref{fig:meta-query}.}
\begin{center}
\begin{tabular}{|c|p{6cm}|p{5cm}|p{4cm}}
Access &SELECT&WHERE&GROUP BY\\\hline
Single &COUNT(*) AS count, $\textless$Attribute-List$\textgreater$, S.sid& $\textless$Key-Equality-List$\textgreater$ AND S.s\_id = jack&  $\textless$Attribute-List$\textgreater$\\
\hline
Block & COUNT(*) AS count,  $\textless$Attribute-List$\textgreater$, S.sid& $\textless$Key-Equality-List$\textgreater$ &  $\textless$Attribute-List$\textgreater$, S.sid\\
\end{tabular}
\end{center}
\label{table:target-query}
\end{table*}%

{\em Assuming} that for each node with ID @parRVID@, a target contingency table named $\CDB.target\_@\it{parRVID}@\_\cttable$ has been built in the Count Database $\CDB$, the log-likelihood SQL is as in Section~\ref{sec:model-score}. For instance, the contribution of the $Capability(\P,\S)$ family is computed by the SQL query shown,  but with the contingency table jack\_Capability(P,S)\_CT in place of Capability(P,S)\_CT.
The new problem is finding the target contingency table. SQL allows us to solve this easily by restricting counts to the target entity in the WHERE clause. To illustrate, suppose we want to modify the contingency table query of Figure~\ref{fig:meta-query} to compute the contingency table for $\S = jack$. We add the student id to the SELECT clause, and the join condition $S.s\_id = jack$ to the WHERE clause; see Table~\ref{table:target-query}.
The FROM clause is the same as in Figure~\ref{fig:meta-query}. The metaquery of Figure~\ref{fig:meta-query} is easily changed to produce these SELECT and WHERE clauses.

Next consider a setting where a model is to be scored against an entire test set. 
For concreteness, suppose the problem is to predict the intelligence of a set of students
 $\it{Intelligence}(jack)$, $\it{Intelligence}(jill)$,
 $\it{Intelligence}(student_{3}),\ldots, \it{Intelligence}(student_{m})$.
 SQL supports {\em block access} where we process the test instances as a block. Intuitively, instead of building a contingency table for each test instance, we build a single contingency table that stacks together the individual contingency tables (Figure~\ref{fig:targetct}). Blocked access can be implemented in a beautifully simple manner in SQL: we simply add the primary key id field for the target entity to the GROUP BY list; see Table~\ref{table:target-query}.

\section{Evaluation} 

Our experimental study describes how \FB\ can be used to implement a challenging machine learning application: Constructing a Bayesian network model for a relational database. Bayesian networks are a good illustration of typical challenges and how RDBMS capabilities can address them because: (1) Bayesian networks are widely regarded as a very useful model class in machine learning and AI, that supports decision making and reasoning under uncertainty. At the same time, they are considered challenging to learn from data. (2) Database researchers have proposed Bayesian networks for combining databases with uncertainty 
\cite{Wang2008}. (3) A Bayesian network with par-RVs can be easily converted to other first-order representations, such as a Markov Logic Network; see \cite{Domingos2009}.

We describe the system and the datasets we used.
Code was written in MySQL Script and Java, JRE 1.7.0.  and executed with 8GB of RAM and a single Intel Core 2 QUAD Processor Q6700 with a clock speed of 2.66GHz (no hyper-threading). The operating system was Linux Centos 2.6.32. 
The MySQL Server version 5.5.34 was run with 8GB of RAM and a single core processor of 2.2GHz. 
All code and datasets are available on-line \cite{bib:bbsite}.

\subsection{Datasets} \label{sec:datasets}
We used six benchmark real-world databases. For detailed descriptions and  the sources of the databases, please see ~\cite{bib:bbsite} and the references therein. Table~\ref{table:datasetsize} summarizes basic information about the benchmark datasets.  
IMDb is the largest dataset in terms of number of total tuples (more than 1.3M tuples) and schema complexity. 
It combines the MovieLens database\footnote{www.grouplens.org, 1M version} with data from the Internet Movie Database (IMDb)\footnote{www.imdb.com, July 2013} following \cite{Peralta2007}.

\begin{table}[hbtp]
\caption{Datasets characteristics. \#Tuples = total number of tuples over all tables in the dataset. }
 \centering
\resizebox{0.5\textwidth}{!}{
\begin{tabular}[c]
{|l|c|c|r|c|}\hline
 \textbf{Dataset} & \textbf{\begin{tabular}[l] {ll} \#Relationship \\Tables/ Total \end {tabular}} & \textbf{\begin{tabular}[l] {ll} \# par-RV \end {tabular}}  & \textbf{\#Tuples} 

\\\hline
    Movielens &1 / 3 & 7  & 1,010,051 
\\\hline
    Mutagenesis & 2 / 4 & 11 & 14,540 
\\\hline
 UW-CSE &2 / 4 & 14  & 712 
\\\hline   
  Mondial &2 / 4 & 18 &  870
\\\hline

   Hepatitis &3 / 7 & 19 &12,927  
\\\hline
   IMDb &3 / 7 & 17 &1,354,134  
\\\hline   
\end{tabular}
}

  \label{table:datasetsize}
\end{table}

Table~\ref{table:datasetsize} provides information about the number of par-RVs generated for each database. More complex schemas 

generate more random variables.

\subsection{Bayesian Network Learning} 
For learning the structure of a first-order Bayesian network, we used \FB\ to implement the previously existing learn-and-join algorithm (LAJ). 
The model search strategy of the LAJ algorithm is an iterative deepening search for correlations among attributes along longer and longer chains of relationships. For more details please see \cite{Schulte2012}. 
The previous implementation of the LAJ algorithm posted at~\cite{bib:bbsite}, limits the par-factors so they contain at most {\em two} relationship par-RVs; \FB\ overcomes this limitation.

A major design decision is how to make sufficient statistics available to the LAJ algorithm. In our experiments we followed a {\em pre-counting} approach where the count manager constructs a \textbf{joint contingency table} for {\em all} par-RVs in the random variable database. An alternative would be {\em on-demand} counting, which computes many contingency tables, but only for factors that are constructed during the model search \cite{Lv2012}.
Pre-counting is a form of data preprocessing: Once the joint contingency table is constructed, local contingency tables can be built quickly by summing (Group By). Different structure learning algorithms can therefore be run quickly on the same joint contingency table. 
For our evaluation, pre-counting has several advantages. (1) Constructing the joint contingency table presents a maximally challenging task for the count manager. (2) Separating counting/data access from model search allows us to assess separately the resources required for each task.

\subsection{Results}
Table~\ref{tab:counts} reports the number of sufficient statistics for constructing the joint contingency table. This number depends mainly on the number of par-RVs. The number of sufficient statistics can be quite large, over 15M for the largest dataset IMDb. 
Even with such large numbers, constructing contingency tables using the SQL metaqueries is feasible, taking just over 2 hours for the very large IMDb set. 
The number of Bayesian network parameters is much smaller than the number of sufficient statistics.
The difference between the number of parameters and the number of sufficient statistics measures how compactly the BN summarizes the statistical information in the data. 
Table~\ref{tab:counts} shows that Bayesian networks provide very compact summaries of the data statistics. For instance for the Hepatitis dataset, the ratio is  $12,374,892/569 > 20,000$. The IMDb database is an outlier, with a complex correlation pattern that leads to a dense Bayesian network structure.

\begin{table}[htbp]
  \caption{Count Manager: Sufficient Statistics and Parameters}
  \centering
   \resizebox{0.5\textwidth}{!}{ \begin{tabular}{|l|r|r|r|r|r|}
    \hline
 \textbf{  Dataset  }&  \textbf{ \begin{tabular}[l] {ll} \# Database\\Tuples  \end{tabular} }&\textbf{ \begin{tabular}[l] {ll}\# Sufficient \\ Statistics (SS) \end{tabular}}& \textbf{\begin{tabular}[l] {ll} SS \\Computing\\ Time (s) \end{tabular}} &\textbf{\begin{tabular}[l] {ll}  \#BN \\ Parameters  \end{tabular}}
\\
    \hline
    Movielens & 1,010,051 & 252   & 2.7   & 292   \\
    \hline
    Mutagenesis & 14,540 & 1,631 & 1.67  & 721   \\
    \hline
    UW-CSE & 712   & 2,828 & 3.84  & 241   \\
    \hline
    Mondial & 870   & 1,746,870 & 1,112.84 & 339  \\
    \hline
    Hepatitis & 12,927 & 12,374,892 & 3,536.76 & 569   \\
    \hline
    IMDb  & 1,354,134 & 15,538,430 & 7,467.85 & 60,059 \\
    \hline
    \end{tabular}%
}
 \label{tab:counts}
\end{table}%

Table~\ref{tab:model} shows that the graph structure of a Bayesian network contains a small number of edges relative to the number of parameters. 
The parameter manager provides fast maximum likelihood estimates for a given structure. 
This is because computing a local contingency table for a BN family is fast given the joint contingency table.
%

\begin{table}[htbp]
\caption{Model Manager Evaluation.}
  \centering
      \resizebox{0.5\textwidth}{!}{ \begin{tabular}{|l|r|r|r|r|}
    \hline
     \textbf{ Dataset} 
&  \textbf{ \begin{tabular}[l] {ll} \# Edges in \\Bayes Net  \end{tabular} }& \textbf{ \begin{tabular}[l] {ll} \# Bayes Net \\ Parameters   \end{tabular} } & \textbf{\begin{tabular}[l] {ll}Parameter \\Learning\\ Time (s) \end{tabular} }\\
    \hline
    Movielens &   72    
    & 292 & 0.57  \\
    \hline
    Mutagenesis &   124   
    & 721 & 0.98  \\
    \hline
    UW-CSE &       112 
   & 241  & 1.14 \\
    \hline
    Mondial &       141 
  & 339 & 60.55    \\
    \hline
    Hepatitis &       207
  & 569  & 429.15   \\
    \hline
    IMDb  &      195 
 & 60,059   & 505.61\\
    \hline
    \end{tabular}%
}  
  \label{tab:model}%
\end{table}%

\begin{figure}[htbp] 
 \centering
\resizebox{0.4\textwidth}{!}{
 \includegraphics[width=0.5\textwidth]{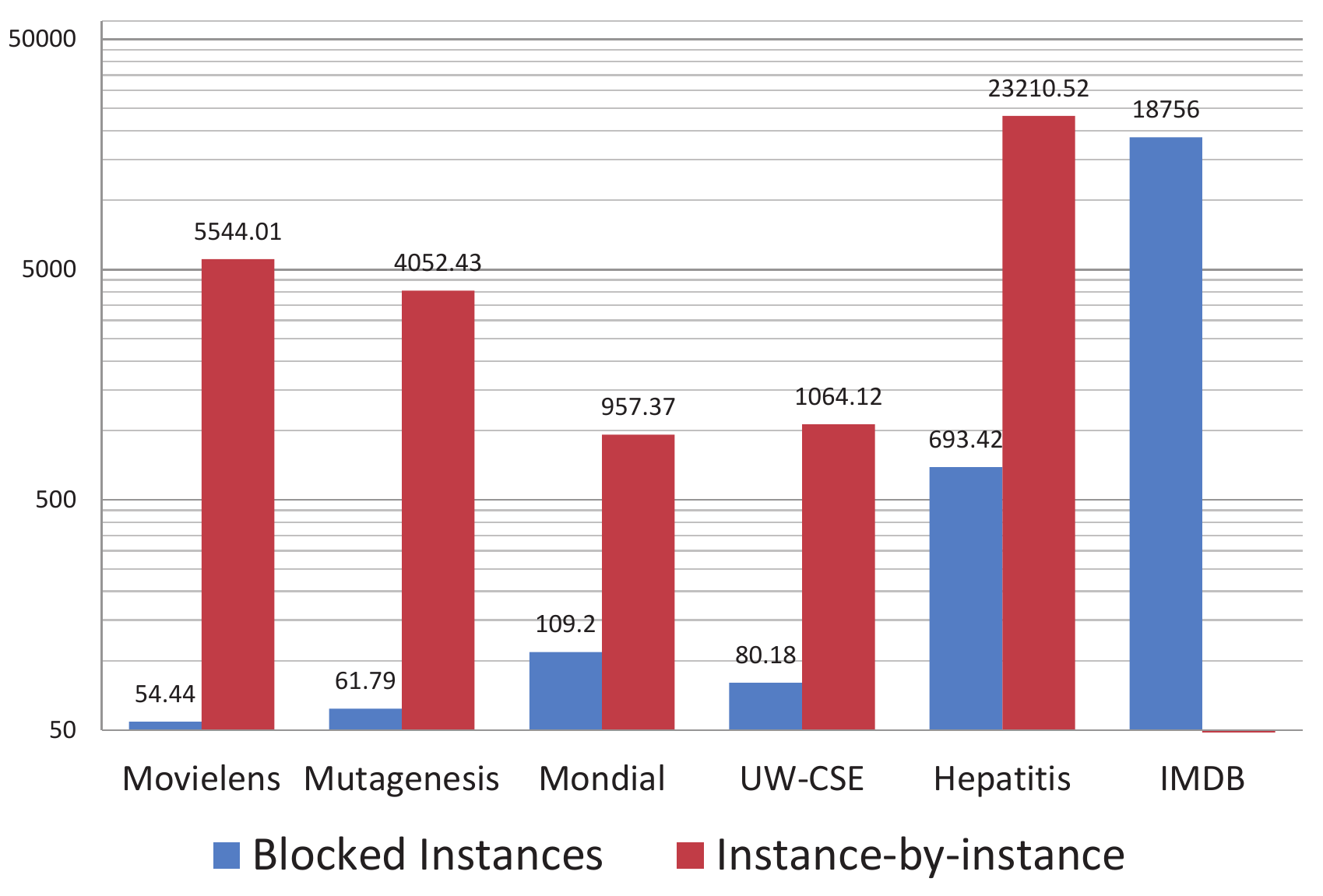} 
} 
\caption{Times (s) for Computing Predictions on Test Instances. The right red column shows the time for looping over single instances using the Single Access Query of Table~\ref{table:target-query}. The left blue column shows the time for the Blocked Access Query of Table~\ref{table:target-query}.
}
 \label{fig:test-timing}
\end{figure}

Figure~\ref{fig:test-timing} compares computing predictions on a test set using an instance-by-instance loop, with a separate SQL query for each instance, vs. a single SQL query for all test instances as a block (Table~\ref{table:target-query}). Table \ref{tab:test-instance} specifies the number of  test instances for each dataset. We split each benchmark database into  80\% training data, 20\% test data. The test instances are the ground atoms of all descriptive attributes of entities.  The blocked access method is 10-100 faster depending on the dataset. The single access method did not scale to the large IMDb dataset (timeout after 12 hours).

\begin{table}[htbp]
\caption{\# of Test Instances }
  \centering
  \begin{tabular}{|r|r|r|r|r|r|r|} \hline
\textbf{Dataset}&Movielens&	Mutagenesis	& 	UW-CSE	&	Mondial&	Hepatitis&	 	IMDb \\ \hline
{\#instance}	&4,742	 	&	3,119		&576	&		505&2,376	 	&46,275 \\ \hline
    
\end{tabular}%
  \label{tab:test-instance}%
\end{table}%

Table~\ref{tab:othersrl} reports result for the complete learning of a Bayesian network, structure and parameters. It benchmarks \FB\ against functional gradient boosting, a state-of-the-art  multi-relational learning approach.
MLN\_Boost learns a Markov Logic Network, and RDN\_Boost a Relational Dependency Network. 
We used the Boostr implementation \cite{Khot2013}. 
To make the results easier to compare across databases and systems, we divide the total running time by the number of par-RVs for the database (Table~\ref{table:datasetsize}). 
Table~\ref{tab:othersrl} shows that structure learning with \FB\ is fast: even the large complex database IMDb requires only around 8 minutes/par-RV. Compared to the boosting methods, \FB\ shows excellent scalability: neither boosting method terminates on the IMDb database, and while RDN\_Boost terminates on the MovieLens database, it is almost 5,000 times slower than {\sc FactorBase}. 
Much of the speed of our implementation is due to quick access to sufficient statistics. As the last column of Table~\ref{tab:othersrl} shows, on the larger datasets \FB\ spends about 80\% of computation time on gathering sufficient statistics via the count manager. This suggests that a large  speedup for the boosting algorithms could be achieved if they used the \FB\ in-database design. 

We do not report accuracy results due to space constraints and because predictive accuracy is not the focus of this paper. On the standard conditional log-likelihood metric, as defined by Equation~\ref{eq:classify}, the model learned by \FB\ performs better than the boosting methods on all databases. This is consistent with the results of previous studies \cite{Schulte2012}.

\begin{table}[htbp] \caption{Learning Time Comparison (sec) with other statistical-relational learning systems. NT = non-termination}
  \centering
      \resizebox{0.45\textwidth}{!}{
\begin{tabular}{|l|r|r|r|r|}\hline
Dataset  & RDN\_Boost  & MLN\_Boost  & FB-Total & FB-Count \\\hline
MovieLens & 5,562  & N/T & 1.12 & 0.39 \\\hline
Mutagenesis  & 118 & 49 & 1 & 0.15 \\\hline
UW-CSE & 15 & 19 & 1 & 0.27 \\\hline
Mondial  & 27 & 42 & 102 & 61.82 \\\hline
Hepatitis  & 251 & 230 & 286 & 186.15 \\\hline
IMDb & N/T & N/T & 524.25 & 439.29 \\\hline
\end{tabular}
} 
  \label{tab:othersrl}%
\end{table}%

{\em Conclusion.} \FB\ leverages RDBMS capabilities for scalable management of statistical analysis objects. It efficiently constructs and stores large numbers of sufficient statistics and parameter estimates. 
The RDBMS support for statistical-relational learning translates into orders of magnitude improvements in speed and scalability.

\section{Related Work} \label{sec:related}

\begin{figure}[htbp] 
 \centering
\resizebox{0.4\textwidth}{!}{
 \includegraphics[width=0.5\textwidth]{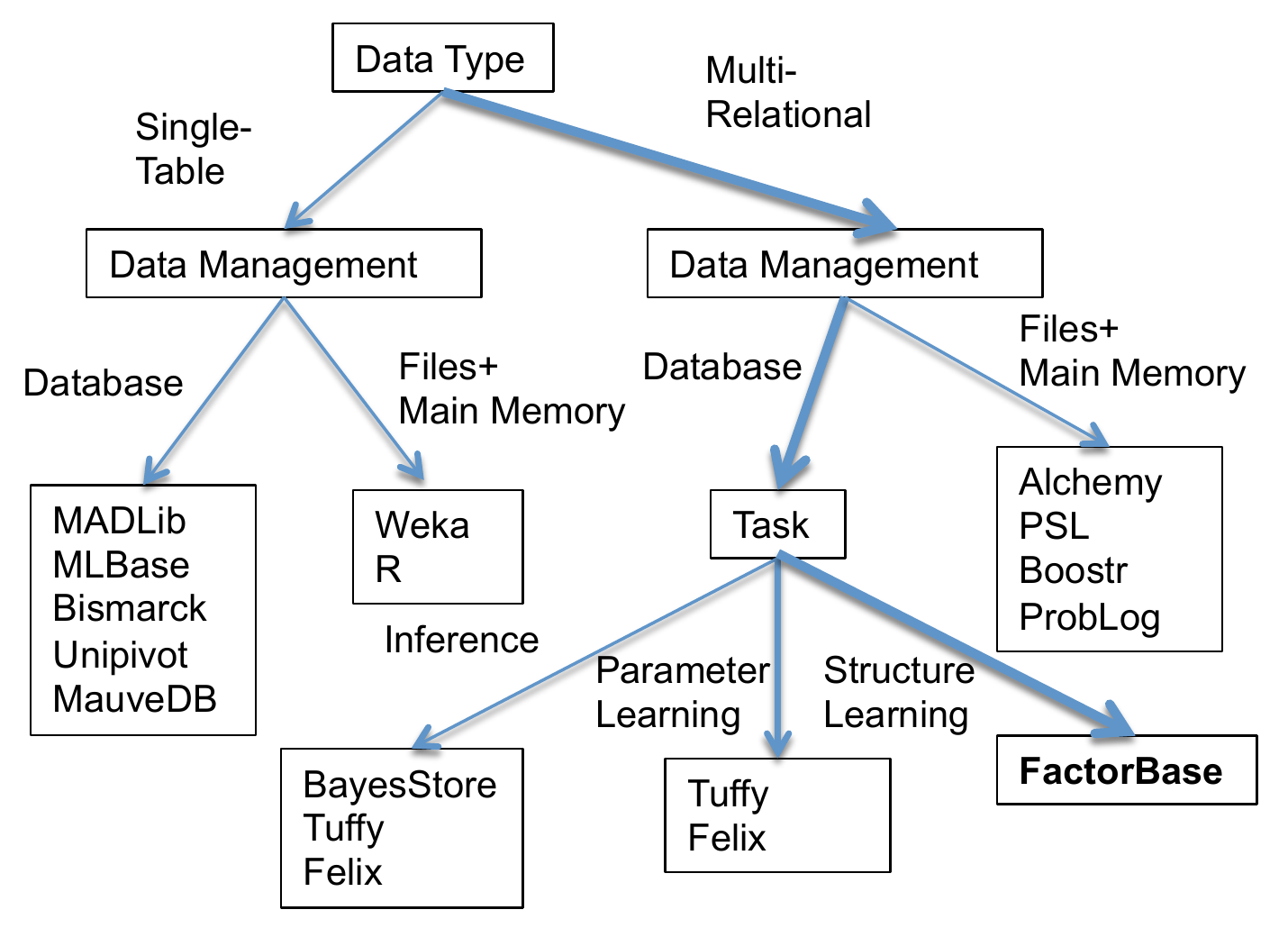} 
 }
\caption{A tree structure for related work in the design space of machine learning $\times$ data management}
\label{fig:related}
\end{figure}

The design space for combining machine learning with data management systems offers a number of possibilities, several of which have been explored in previous and ongoing research. 
We selectively review the work most relevant to our research. Figure~\ref{fig:related} provides a tree structure for the research landscape. 

\subsection{Single-Table Machine Learning} Most machine learning systems, such as Weka or R, support learning from a single table or data matrix only. The single-table representation is appropriate when the data points represent a homogeneous class of entities with similar attributes, where the attributes of one entity are independent of those of others \cite{Kimmig2015}. The only way a single-table system can be applied to multi-relational data is after a preprocessing step where multiple interrelated tables are converted to a single data table. When the learning task is classification, such preprocessing is often called propositionalization  \cite{Kimmig2015}.  This ``flattening'' of the relational structure typically involves a loss of information.  

\subsubsection{RDBMS Learning}
Leveraging RDBMS capabilities through SQL programming 
is the unifying idea of the recent MADLib framework \cite{MADlib_VLDB_2012}. An advantage of the MADLib approach that is shared by \FB\ is that in-database processing avoids exporting the data from the input database. The Apache Spark \cite{Committers} framework includes MLBase and SparkSQL that provide support for distributed processing, SQL, and automatic refinement of machine learning algorithms and models~\cite{MLbase_ICDR_2013}.
Other RDBMS applications include gathering sufficient statistics \cite{Graefe1998}, and convex optimization \cite{Feng_SIGMOD_2012}. The MauveDB system \cite{Deshpande2006} emphasizes the importance of several RDBMS features for combining statistical analysis with databases.
As in {\sc FactorBase}, this includes 
storing models and associated parameters as objects in their own right, 
and using the view mechanism to update statistical objects as the data change.
A difference is that
MauveDB presents model-based views of the {\em data} to the user, whereas \FB\ presents views of the {\em models} to machine learning applications. 

\subsubsection{RDBMS Inference}
Wong {\em et al.}  applied SQL operators such as the natural join to perform log-linear inference with a single-table graphical model \cite{Wong1995} stored in an RDBMS. 
Monte Carlo methods have also been implemented with an RDBMS  to perform inference with uncertain data~\cite{MCDB_SIGMOD_2008,Wick_VLDB_2010}.
The MCDB system \cite{MCDB_SIGMOD_2008}  stores parameters in database tables like \FB.

\subsection{Multi-Relational Learning} 
For overviews of multi-relational learning please see \cite{SRL2007,Domingos2009,Kimmig2015}. 
Most implemented systems, such as Aleph and Alchemy, use a logic-based representation of data derived from Prolog facts, that originated in the Inductive Logic Programming community \cite{Dzeroski2001c}. 

\subsubsection{RDBMS Learning}
The ClowdFlows system \cite{Lavravc2014} allows a user to specify a MySQL database as a data source, then converts the MySQL data to a single-table representation using propositionalization. 
Singh and Graepel \cite{Singh2013} present an algorithm that analyzes the relational database system catalog to generate a set of nodes and a Bayesian network structure. 
This approach utilizes SQL constructs as a data description language in a way that is similar to our Schema Analyzer. 
Differences include the following. (1) The Bayesian network structure is fixed and based on latent variables, rather than learned for observable variables only, as in our case study. (2) The RDBMS is not used to support learning after random variables have been extracted from the schema.

Qian {\em et al.} \cite{Qian2014a}  discuss work related to the contingency table problem and introduce contingency table algebra. Their paper focuses on a Virtual Join algorithm for computing sufficient statistics that involve negated relationships. They do not discuss integrating contingency tables with other structured objects for multi-relational learning.

\subsubsection{RDBMS Inference} 
Database researchers have developed powerful probabilistic inference  algorithms for multi-relational models. 
 The BayesStore system \cite{Wang2008} introduced the principle of treating all statistical objects as first-class citizens in a relational database as \FB\ does. The Tuffy system \cite{Niu2011} achieves highly reliable and scalable inference for Markov Logic Networks (MLNs) with an RDBMS. It leverages inference capabilities to perform MLN parameter learning. RDBMS support for local search parameter estimation procedures, rather than closed-form maximum-likelihood estimation, has also been explored \cite{Feng_SIGMOD_2012,Niu2011,Niu2011a}.

\section{Conclusion and Future Work}

Compared to traditional learning with a single data table, learning for multi-relational data requires new system capabilities. In this paper we described \FB, a system that leverages the existing capabilities of an SQL-based RDBMS to support statistical-relational learning.
Representational tasks include specifying metadata about structured first-order random variables, and storing the structure of a learned model. Computational tasks include storing and constructing sufficient statistics, and computing parameter estimates and model selection scores. 
We showed that SQL scripts can be used to implement these capabilities, with multiple advantages. These advantages include: 1) Fast program development through high-level SQL constructs for complex table and count operations. 2) Managing large and complex statistical objects that are too big to fit in main memory. 
For instance, some of our benchmark databases require storing and querying millions of sufficient statistics. While \FB\ provides good solutions for each of these system capabilities in isolation, the ease with which large complex statistical-relational objects can be integrated via SQL queries is a key feature. 
Because information about random variables, sufficient statistics, and models 
is all represented in relational database tables,
a machine learning application can access and combine the information in a uniform way via SQL queries.

Empirical evaluation on six benchmark databases showed significant scalability advantages from utilizing the RDBMS capabilities: Both structure and parameter learning scaled well to millions of data records, beyond what previous multi-relational learning systems can achieve.

{\em Future Work.} Further potential application areas for \FB\ include managing massive numbers of aggregate features for classification \cite{Popescul2007}, and collective matrix factorization \cite{Singh2008,Singh2013}. 
While our implementation has used simple SQL plus indexes, there are opportunities for optimizing RDBMS operations for the workloads required by statistical-relational structure learning. These include view materialization and the key scalability bottleneck of computing multi-relational sufficient statistics. NoSQL databases can exploit a flexible data representation for scaling to very large datasets. However, SRL requires count operations for random complex join queries, which is a challenge for less structured data representations. An important goal is a single RDBMS package for both learning and inference that integrates \FB\ with inference systems such as BayesStore and Tuffy. 
There are several fundamental system design choices whose trade-offs for SRL warrant exploration. These include the choice between pre-counting and post-counting sufficient statistics, and using main memory vs. RDBMS disk storage. For instance, model selection scores can be cached in either main memory or the database. Our SQL-based approach facilitates using distributed computing systems such as SparkSQL \cite{Michael2015}, which have shown great potential for scalability. 
In sum, we believe that the succesful use of SQL presented in this paper shows that relational algebra can play the same role for multi-relational learning as linear algebra for single-table learning: a unified language for both representing statistical objects and for computing with them.

\section*{Acknowledgement} 
This research was supported by a Discovery grant to Oliver Schulte by the Natural Sciences and Engineering Research Council of Canada. 
Zhensong Qian was supported by a grant from the China Scholarship Council.
A position paper based on this work was presented at the StarAI 2015 workshop. We thank the workshop reviewers and participants for helpful comments.

\bibliographystyle{abbrv}
\bibliography{pre-print} 

\begin{appendix} \label{sec:appendix}

\section{The Random Variable Database Layout}
\begin{figure}[htbp]
\begin{center}
\resizebox{0.5\textwidth}{!}{
\includegraphics[width=0.5\textwidth]{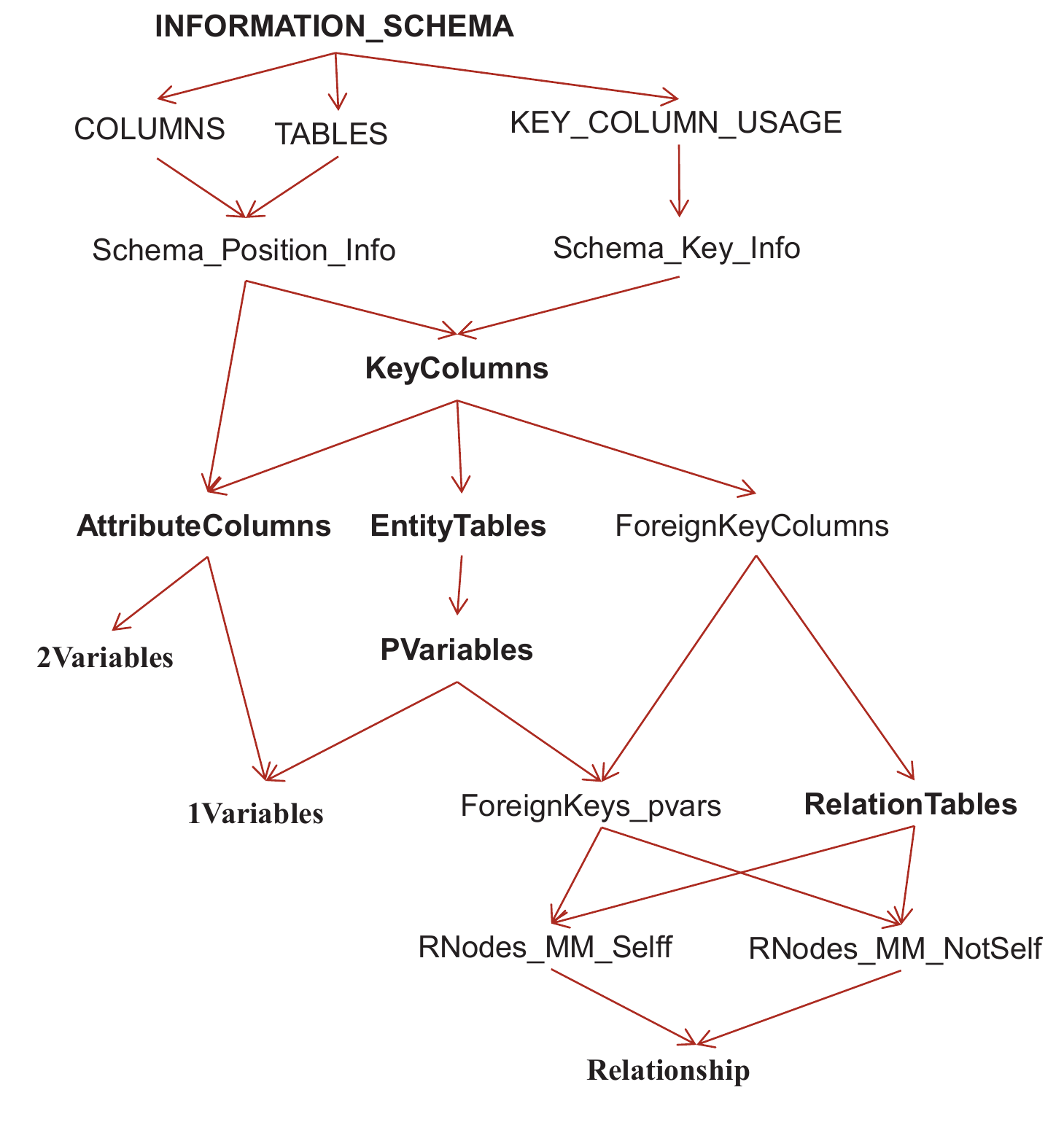} 
}
\caption{Tables  Dependency in the Random Variable Database $\RVD$.
\label{fig:rv_db1}}
\end{center}
\end{figure}

We provide details about the Schema Analyzer. Table~\ref{table:rvdb1} shows the relational schema of the Random Variable Database. Figure~\ref{fig:rv_db1} shows dependencies between the tables of this schema. 

\begin{table}[htbp]
  \centering
  \caption{Schema for Random Variable Database}
\resizebox{0.5\textwidth}{!}{
    \begin{tabular}{|r|r|r|r|r|r|}
    \hline
    \multicolumn{2}{|c|}{Table Name} & \multicolumn{4}{c|}{Schema}  \\
    \hline
    \multicolumn{2}{|l|}{AttributeColumns} & \multicolumn{4}{l|}{\begin{tabular}{l}TABLE\_NAME,  COLUMN\_NAME   \end{tabular}}  \\
    \hline
    \multicolumn{2}{|l|}{Domain} & \multicolumn{4}{l|}{\begin{tabular}{l}COLUMN\_NAME, VALUE   \end{tabular}}  \\
    \hline
    \multicolumn{2}{|l|}{Pvariables} & \multicolumn{4}{l|}{\begin{tabular}{l}Pvid,  TABLE\_NAME  \end{tabular}}  \\
    \hline
    \multicolumn{2}{|l|}{1Variables} & \multicolumn{4}{l|}{\begin{tabular}{l}1VarID,  COLUMN\_NAME,  Pvid 
\end{tabular} }  \\
    \hline
    \multicolumn{2}{|l|}{2Variables } & \multicolumn{4}{l|}{\begin{tabular}{ll} 2VarID,  COLUMN\_NAME,  Pvid1,  Pvid2, \\ TABLE\_NAME \end{tabular}}  \\
    \hline
    \multicolumn{2}{|l|}{Relationship} & \multicolumn{4}{l|}{\begin{tabular}{lll}RVarID,  TABLE\_NAME, Pvid1,  Pvid2, \\ COLUMN\_NAME1,  COLUMN\_NAME2  \end{tabular}}  \\
    \hline
    \end{tabular}%
}

  \label{table:rvdb1}%
\end{table}%

\section{MySQL Script for creating default random variables.}

\begin{scriptsize}
\begin{alltt}
/*AchemaAnalyzer.sql*/
DROP SCHEMA IF EXISTS @database@_AchemaAnalyzer; 
CREATE SCHEMA  @database@_AchemaAnalyzer;

CREATE SCHEMA  if not exists @database@_BN;
CREATE SCHEMA  if not exists @database@_CT;

USE @database@_AchemaAnalyzer;
SET storage_engine=INNODB;

CREATE TABLE Schema_Key_Info AS SELECT TABLE_NAME, COLUMN_NAME,
REFERENCED_TABLE_NAME, REFERENCED_COLUMN_NAME, CONSTRAINT_NAME 
FROM INFORMATION_SCHEMA.KEY_COLUMN_USAGE 
WHERE (KEY_COLUMN_USAGE.TABLE_SCHEMA =
'@database@') ORDER BY TABLE_NAME;

CREATE TABLE Schema_Position_Info AS 
SELECT COLUMNS.TABLE_NAME,
COLUMNS.COLUMN_NAME,
COLUMNS.ORDINAL_POSITION FROM
INFORMATION_SCHEMA.COLUMNS,
INFORMATION_SCHEMA.TABLES
WHERE
(COLUMNS.TABLE_SCHEMA = '@database@'
    AND TABLES.TABLE_SCHEMA = '@database@'
    AND TABLES.TABLE_NAME = COLUMNS.TABLE_NAME
    AND TABLES.TABLE_TYPE = 'BASE TABLE')
ORDER BY TABLE_NAME;

CREATE TABLE NoPKeys AS SELECT TABLE_NAME FROM
Schema_Key_Info
WHERE
TABLE_NAME NOT IN (SELECT 
        TABLE_NAME
    FROM
        Schema_Key_Info
    WHERE
        CONSTRAINT_NAME LIKE 'PRIMARY');

CREATE table NumEntityColumns AS
SELECT 
    TABLE_NAME, COUNT(DISTINCT COLUMN_NAME) num
FROM
    Schema_Key_Info
WHERE
    CONSTRAINT_NAME LIKE 'PRIMARY'
        OR REFERENCED_COLUMN_NAME IS NOT NULL
GROUP BY TABLE_NAME;

CREATE TABLE TernaryRelations as SELECT TABLE_NAME FROM
NumEntityColumns
WHERE
num > 2;


CREATE TABLE AttributeColumns AS
SELECT TABLE_NAME, COLUMN_NAME FROM
Schema_Position_Info
WHERE
(TABLE_NAME , COLUMN_NAME) NOT IN (SELECT 
        TABLE_NAME, COLUMN_NAME
    FROM
        KeyColumns)
    and TABLE_NAME NOT IN (SELECT 
        TABLE_NAME
    FROM
        NoPKeys)
    and TABLE_NAME NOT IN (SELECT 
        TABLE_NAME
    FROM
        TernaryRelations);

ALTER TABLE AttributeColumns
ADD PRIMARY KEY (TABLE_NAME,COLUMN_NAME);

CREATE TABLE InputColumns AS SELECT * FROM
KeyColumns
WHERE
CONSTRAINT_NAME = 'PRIMARY'
ORDER BY TABLE_NAME;

CREATE TABLE ForeignKeyColumns AS SELECT * FROM
KeyColumns
WHERE
REFERENCED_COLUMN_NAME IS NOT NULL
ORDER BY TABLE_NAME;

ALTER TABLE ForeignKeyColumns 
ADD PRIMARY KEY (TABLE_NAME,COLUMN_NAME,
REFERENCED_TABLE_NAME);

CREATE TABLE EntityTables AS 
SELECT distinct TABLE_NAME, COLUMN_NAME 
FROM
KeyColumns T
WHERE
1 = (SELECT 
        COUNT(COLUMN_NAME)
    FROM
        KeyColumns T2
    WHERE
        T.TABLE_NAME = T2.TABLE_NAME
            AND CONSTRAINT_NAME = 'PRIMARY');

ALTER TABLE EntityTables 
ADD PRIMARY KEY (TABLE_NAME,COLUMN_NAME);

CREATE TABLE SelfRelationships AS 
SELECT DISTINCT RTables1.TABLE_NAME 
AS TABLE_NAME,
RTables1.REFERENCED_TABLE_NAME AS REFERENCED_TABLE_NAME,
RTables1.REFERENCED_COLUMN_NAME AS REFERENCED_COLUMN_NAME FROM
KeyColumns AS RTables1,
KeyColumns AS RTables2
WHERE
(RTables1.TABLE_NAME = RTables2.TABLE_NAME)    AND 
(RTables1.REFERENCED_TABLE_NAME = RTables2.REFERENCED_TABLE_NAME)
 AND 
(RTables1.REFERENCED_COLUMN_NAME = RTables2.REFERENCED_COLUMN_NAME)
 AND 
(RTables1.ORDINAL_POSITION < RTables2.ORDINAL_POSITION);

ALTER TABLE SelfRelationships ADD PRIMARY KEY (TABLE_NAME);

CREATE TABLE Many_OneRelationships AS 
SELECT KeyColumns1.TABLE_NAME FROM
KeyColumns AS KeyColumns1,
KeyColumns AS KeyColumns2
WHERE
(KeyColumns1.TABLE_NAME , KeyColumns1.COLUMN_NAME) IN (SELECT 
        TABLE_NAME, COLUMN_NAME
    FROM         InputColumns)
    AND (KeyColumns2.TABLE_NAME , KeyColumns2.COLUMN_NAME) IN 
  (SELECT  TABLE_NAME, COLUMN_NAME
    FROM    ForeignKeyColumns)
    AND (KeyColumns2.TABLE_NAME , KeyColumns2.COLUMN_NAME)
   NOT IN (SELECT  TABLE_NAME, COLUMN_NAME
    FROM     InputColumns);

CREATE TABLE PVariables AS 
SELECT CONCAT(EntityTables.TABLE_NAME, '0') AS Pvid,
EntityTables.TABLE_NAME,
0 AS index_number FROM
EntityTables 
UNION 
SELECT 
CONCAT(EntityTables.TABLE_NAME, '1') AS Pvid,
EntityTables.TABLE_NAME,
1 AS index_number
FROM
EntityTables,
SelfRelationships
WHERE
EntityTables.TABLE_NAME = SelfRelationships.REFERENCED_TABLE_NAME
AND 
EntityTables.COLUMN_NAME = SelfRelationships.REFERENCED_COLUMN_NAME;

ALTER TABLE PVariables ADD PRIMARY KEY (Pvid);

CREATE TABLE RelationTables AS 
SELECT DISTINCT ForeignKeyColumns.TABLE_NAME,
ForeignKeyColumns.TABLE_NAME IN (SELECT 
        TABLE_NAME
    FROM
        SelfRelationships) AS SelfRelationship,
ForeignKeyColumns.TABLE_NAME IN (SELECT 
        TABLE_NAME
    FROM
        Many_OneRelationships) AS Many_OneRelationship FROM
ForeignKeyColumns;

ALTER TABLE RelationTables 
ADD PRIMARY KEY (TABLE_NAME);

CREATE TABLE 1Variables AS 
SELECT CONCAT('`', COLUMN_NAME, '(', Pvid, ')', '`') AS 1VarID,
COLUMN_NAME,
Pvid,
index_number = 0 AS main FROM
PVariables
    NATURAL JOIN
AttributeColumns;

ALTER TABLE 1Variables ADD PRIMARY KEY (1VarID);
ALTER TABLE 1Variables ADD UNIQUE(Pvid,COLUMN_NAME);

CREATE TABLE ForeignKeys_pvars AS 
SELECT ForeignKeyColumns.TABLE_NAME,
ForeignKeyColumns.REFERENCED_TABLE_NAME,
ForeignKeyColumns.COLUMN_NAME,
Pvid,
index_number,
ORDINAL_POSITION AS ARGUMENT_POSITION FROM
ForeignKeyColumns,
PVariables
WHERE
PVariables.TABLE_NAME = REFERENCED_TABLE_NAME;

ALTER TABLE ForeignKeys_pvars 
ADD PRIMARY KEY (TABLE_NAME,Pvid,
ARGUMENT_POSITION);

CREATE table Relationship_MM_NotSelf AS
SELECT 
    CONCAT('`',
            ForeignKeys_pvars1.TABLE_NAME,
            '(',
            ForeignKeys_pvars1.Pvid,
            ',',
            ForeignKeys_pvars2.Pvid,
            ')',
            '`') AS orig_RVarID,
    ForeignKeys_pvars1.TABLE_NAME,
    ForeignKeys_pvars1.Pvid AS Pvid1,
    ForeignKeys_pvars2.Pvid AS Pvid2,
    ForeignKeys_pvars1.COLUMN_NAME AS COLUMN_NAME1,
    ForeignKeys_pvars2.COLUMN_NAME AS COLUMN_NAME2,
    (ForeignKeys_pvars1.index_number = 0
        AND ForeignKeys_pvars2.index_number = 0) AS main
FROM
    ForeignKeys_pvars AS ForeignKeys_pvars1,
    ForeignKeys_pvars AS ForeignKeys_pvars2,
    RelationTables
WHERE
 ForeignKeys_pvars1.TABLE_NAME = ForeignKeys_pvars2.TABLE_NAME
 AND RelationTables.TABLE_NAME = ForeignKeys_pvars1.TABLE_NAME
        AND ForeignKeys_pvars1.ARGUMENT_POSITION <
        ForeignKeys_pvars2.ARGUMENT_POSITION
        AND RelationTables.SelfRelationship = 0
        AND RelationTables.Many_OneRelationship = 0;

CREATE table Relationship_MM_Self AS
SELECT 
    CONCAT('`',
            ForeignKeys_pvars1.TABLE_NAME,
            '(',
            ForeignKeys_pvars1.Pvid,
            ',',
            ForeignKeys_pvars2.Pvid,
            ')',
            '`') AS orig_RVarID,
    ForeignKeys_pvars1.TABLE_NAME,
    ForeignKeys_pvars1.Pvid AS Pvid1,
    ForeignKeys_pvars2.Pvid AS Pvid2,
    ForeignKeys_pvars1.COLUMN_NAME AS COLUMN_NAME1,
    ForeignKeys_pvars2.COLUMN_NAME AS COLUMN_NAME2,
    (ForeignKeys_pvars1.index_number = 0
        AND ForeignKeys_pvars2.index_number = 1) AS main
FROM
    ForeignKeys_pvars AS ForeignKeys_pvars1,
    ForeignKeys_pvars AS ForeignKeys_pvars2,
    RelationTables
WHERE
ForeignKeys_pvars1.TABLE_NAME = ForeignKeys_pvars2.TABLE_NAME
AND RelationTables.TABLE_NAME = ForeignKeys_pvars1.TABLE_NAME
AND ForeignKeys_pvars1.ARGUMENT_POSITION <
ForeignKeys_pvars2.ARGUMENT_POSITION
AND 
ForeignKeys_pvars1.index_number < ForeignKeys_pvars2.index_number
AND RelationTables.SelfRelationship = 1
AND RelationTables.Many_OneRelationship = 0;

CREATE table Relationship_MO_NotSelf AS
SELECT 
    CONCAT('`',
            ForeignKeys_pvars.REFERENCED_TABLE_NAME,
            '(',
            PVariables.Pvid,
            ')=',
            ForeignKeys_pvars.Pvid,
            '`') AS orig_RVarID,
    ForeignKeys_pvars.TABLE_NAME,
    PVariables.Pvid AS Pvid1,
    ForeignKeys_pvars.Pvid AS Pvid2,
    KeyColumns.COLUMN_NAME AS COLUMN_NAME1,
    ForeignKeys_pvars.COLUMN_NAME AS COLUMN_NAME2,
    (PVariables.index_number = 0
        AND ForeignKeys_pvars.index_number = 0) AS main
FROM
    ForeignKeys_pvars,
    RelationTables,
    KeyColumns,
    PVariables
WHERE
    RelationTables.TABLE_NAME = ForeignKeys_pvars.TABLE_NAME
        AND RelationTables.TABLE_NAME = PVariables.TABLE_NAME
        AND RelationTables.TABLE_NAME = KeyColumns.TABLE_NAME
        AND RelationTables.SelfRelationship = 0
        AND RelationTables.Many_OneRelationship = 1;

CREATE table Relationship_MO_Self AS
SELECT 
    CONCAT('`',
            ForeignKeys_pvars.REFERENCED_TABLE_NAME,
            '(',
            PVariables.Pvid,
            ')=',
            ForeignKeys_pvars.Pvid,
            '`') AS orig_RVarID,
    ForeignKeys_pvars.TABLE_NAME,
    PVariables.Pvid AS Pvid1,
    ForeignKeys_pvars.Pvid AS Pvid2,
    KeyColumns.COLUMN_NAME AS COLUMN_NAME1,
    ForeignKeys_pvars.COLUMN_NAME AS COLUMN_NAME2,
    (PVariables.index_number = 0
        AND ForeignKeys_pvars.index_number = 1) AS main
FROM
    ForeignKeys_pvars,
    RelationTables,
    KeyColumns,
    PVariables
WHERE
    RelationTables.TABLE_NAME = ForeignKeys_pvars.TABLE_NAME
        AND RelationTables.TABLE_NAME = PVariables.TABLE_NAME
        AND RelationTables.TABLE_NAME = KeyColumns.TABLE_NAME
AND PVariables.index_number < ForeignKeys_pvars.index_number
        AND RelationTables.SelfRelationship = 1
        AND RelationTables.Many_OneRelationship = 1;

CREATE TABLE Relationship AS SELECT * FROM  
Relationship_MM_NotSelf    
UNION SELECT            
*                   
FROM
Relationship_MM_Self 
UNION SELECT 
*
FROM
Relationship_MO_NotSelf 
UNION SELECT 
*
FROM
Relationship_MO_Self;

ALTER TABLE Relationship ADD PRIMARY KEY (orig_RVarID);
ALTER TABLE `Relationship` 
ADD COLUMN `RVarID` VARCHAR(10) NULL , 
ADD UNIQUE INDEX `RVarID_UNIQUE` (`RVarID` ASC) ; 


CREATE TABLE 2Variables AS SELECT CONCAT('`',
        COLUMN_NAME,
        '(',
        Pvid1,
        ',',
        Pvid2,
        ')',
        '`') AS 2VarID,
COLUMN_NAME,
Pvid1,
Pvid2,
TABLE_NAME,
main FROM
Relationship    NATURAL JOIN  AttributeColumns;

ALTER TABLE 2Variables ADD PRIMARY KEY (2VarID);</p>\end{alltt}

\end{scriptsize}

\end{appendix}

\end{document}